\documentclass{article}
\usepackage{spconf,amsmath,graphicx}
\usepackage{times}
\usepackage{amssymb}
\usepackage{graphicx}
\usepackage{float}
\usepackage{url}
\usepackage{amsmath}
\usepackage{subfig}
\usepackage{algorithmic}
\newcommand{\mycomment}[1]{}

\title{Robust Phonetic Segmentation Using Spectral Transition Measure for Non-standard Recording Environments}
\name{Bhavik Vachhani, Chitralekha Bhat, Sunil Kopparapu}
\address{TCS Innovation Labs, Mumbai, India\\
	  Email: bhavik.vachhani@tcs.com, bhat.chitralekha@tcs.com, sunilkumar.kopparapu@tcs.com}
\begin{document}
\maketitle
\begin{abstract}
Phone level localization of mis-articulation 
is a key requirement for an
automatic articulation error assessment system.
A robust phone segmentation technique is essential 
to aid in real-time assessment of phone level mis-articulations 
of speech, wherein the audio is recorded on mobile phones or tablets. 
This is a non-standard recording set-up with little control 
over the quality of  recording.
We propose a novel 
post processing technique 
to aid Spectral Transition Measure(STM)-based
phone segmentation under noisy conditions such as 
environment noise and clipping, commonly present
during a mobile phone recording. A comparison 
of the performance of our approach 
and phone segmentation using traditional 
MFCC and PLPCC speech features 
for Gaussian noise and 
clipping is shown. The proposed approach was 
validated on TIMIT and Hindi speech corpus
and was used to compute phone boundaries 
for a set of speech, recorded simultaneously 
on three devices - a laptop, a stationarily placed tablet 
and a handheld mobile phone, to simulate
different audio qualities in a real-time non-standard
recording environment.
F-ratio was the metric used to compute the
accuracy in phone boundary marking. Experimental
results show an improvement of 7\% for TIMIT
and 10\% for Hindi data over the baseline approach.
Similar results were seen for the set of three of recordings
collected in-house.
\end{abstract}
\section{Introduction}
\label{sec:intro}
Phonetic segmentation is the process of breaking down
a given speech utterance into its basic units, namely phones.
Accurate and robust phonetic segmentation is a key 
requirement for an automatic mis-articulation assessment system, 
wherein the spoken speech could be from a patient 
undergoing speech language therapy or a student wishing to 
learn a new language. Feedback needs to be given based on the 
accuracy of pronunciation of each phone, thus making 
phone level localization of mis-articulation essential.
Several techniques have been proposed for 
phonetic segmentation, popular ones being, 
Automatic Speech Recognition (ASR) based, 
wavelet analysis based and Spectral Transition Measure (STM) based.
In \cite{YuanRLSMW13,Adell04towardsphone,PatilJR09} authors, 
address the accuracy of the phonetic segmentation 
using a two-step approach, wherein the initial estimate is obtained
using an ASR and the boundaries are further refined using specific
boundary level acoustic models, approach based on regression tree
and on acoustic-phonetic knowledge, respectively. Authors use similar
approach in \cite{Stolcke}, wherein the boundaries were improved using powerful
statistical models conditioned on phonetic context 
and duration features. However, any ASR based
method is dependent on the availability and quality of speech corpus
 in a particular language. 
Phonetic boundaries have been computed using 
several speech parameterizations such as 
Mel Frequency Cepstral Coefficients (MFCC),
Perceptual Linear Prediction Cepstral Coefficients (PLPCC), 
RelAtive SpecTrAl (RASTA)-based PLPCC, wavelet-based parameters.
 In \cite{Mporas} authors use Fourier-based and wavelet-based parameterization 
for a Viterbi time-alignment based phonetic segmentation. A phonetic
segmentation algorithm based on power fluctuations of the wavelet
spectrum for a speech signal has been proposed in \cite{Ziolko}.
Discrete Wavelet transform (DWT), its power spectrum and its derivatives
have been used to achieve phonetic segmentation. Phonetic segmentation
and classification using wavelet based transforms was used
to enhance frequencies adaptively in hearing aids by 
authors in \cite{Tan94applyingwavelet}.
Maximum marginal clustering (MMC), a kernel method has
been applied for unsupervised phonetic segmentation \cite{Estevan}.
Phonetic boundaries are detected using a two-layered support vector
machine (SVM)-based system using frequency synchrony and average signal
levels computed using a biomimetic model of 
the human auditory processing \cite{King}. 
These techniques have been reported to perform 
well on clean speech.   

However, speech based mobile applications such as Siri,
are gaining popularity and we envision that
a robust mechanism for phonetic segmentation
of recordings conducted in such non-standard set-ups
are valuable to building speech based applications on 
mobile phones. Specifically, our objective
is to assess mis-articulations in disordered 
speech at phone level and provide instant feedback
to the user for corrective action through hand-held
devices such as mobile phones or tablets. 
However, mobile phone or tablet recordings 
are perpetuated with
environment noise and clipping. 
Phonetic segmentation using traditional methods and speech parameters 
alone will not provide the levels of accuracy required
for such a task.  
Spectral Transition Measure is closely correlated with phonetic boundaries \cite{Sdusan} 
and hence can be exploited to automatically obtain phonetic boundaries
in a language independent manner. STM based methods
have also been recently used to analyze the 
effectiveness of Perceptual linear prediction
(PLP) based features in speech synthesis \cite{Bhavik}.
We propose a novel post processing  
mechanism for Spectral Transition Measure (STM) based 
phonetic segmentation, through which accuracies
for noisy speech (environmental noise and clipping) was improved, which is the main contribution
of this paper.

The organization of the paper is as follows:
Section \ref{sec:STM} describes the STM-based phonetic segmentation approach
and its limitations . 
Section \ref{sec:Distortions} discusses the distortions in
speech recorded using handheld devices such as mobile phones and tablets.
Section \ref{sec:Experimental setup} describes the design of
our experiments for evaluating the proposed technique.
Section \ref{sec:Evaluation} discusses the results from discusses the evaluation results.
Finally,Section \ref{sec:conclusion} concludes the paper along with directions for future work.

\section{Phonetic segmentation algorithm}
\label{sec:STM}
\begin{figure}[H]
\includegraphics[width=0.470\textwidth,height=0.30\textheight]{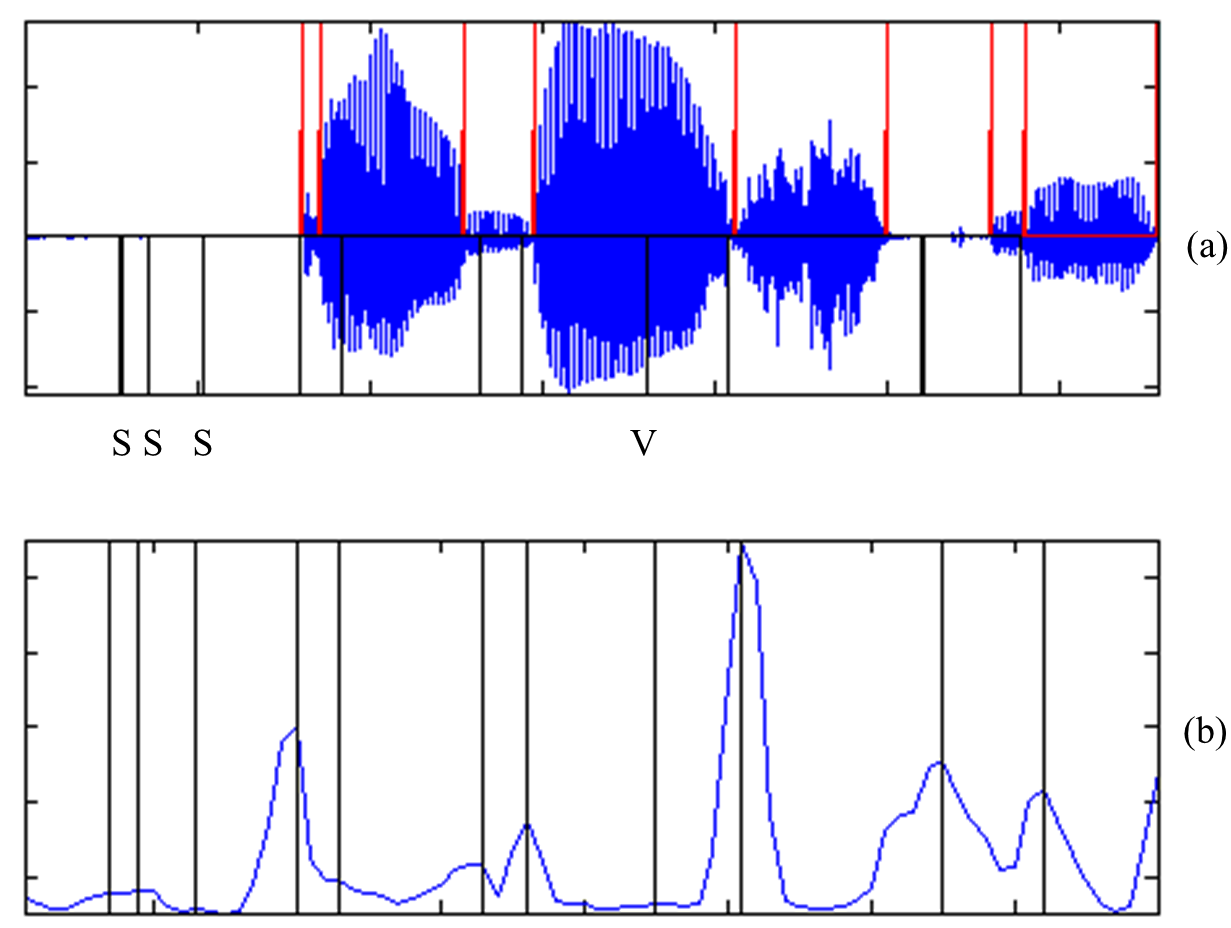}
\caption{{(a) Original signal (\emph{'Don't Ask Me'} from TIMIT) with corresponding manually marked boundaries 
\emph{(red lines)} and automatic segmented boundaries \emph{(black lines)} (b) STM contour using baseline algorithm.}}  
\label{figure:original_spectrogram}
\end{figure}

In literature various automatic phonetic segmentation algorithms are reported. However, in our work we have used STM algorithm 
for automatic phonetic segmentation \cite{SFurai}.
For a given speech signal we compute STM is as follows,
\begin{displaymath}
{f} = [\bar{f_1},\bar{f_2},...,\bar{f_m}]\\  
\end{displaymath}
where $\bar{f}_i$ is spectral feature vectors of D-dimension and \emph{m} is total number of frames for a given speech signal. 
Than, Equation \ref{eq:1} defined the rate of spectral feature with $\bar{a} = [a_1, a_2,...,a_D]$.
\begin{equation} 
\bar{a}(m) = \frac{\sum_{n=-I}^{I}\bar{f}_{(n+m)}.n}{\sum_{n=-I}^{I}n^2}
\label{eq:2}
\end{equation}
\begin{equation} 
STM(m) = \frac{\sum_{}^{}\bar{a}(m)^2}{D}
\label{eq:1}
\end{equation}\\
STM is defined as mean-squared value of spectral rate using Equation \ref{eq:1}.
%
The locations of local maxima obtained 
in the STM contour indicate
spectral transition or a phone boundary.
The STM algorithm for phonetic segmentation 
was originally proposed using 
MFCC features \cite{SFurai,Sdusan}.
In \cite{Bhavik}, authors have achieved improvement 
over the state-of-the-art using PLPCC features for 
automatic phonetic segmentation. In our work we are using PLPCC base STM as our baseline algorithm.
Figure \ref{figure:original_spectrogram} depicts the STM contour
corresponding to a clean speech utterance wherein
phone boundaries are captured using peak-picking.
\subsection{Limitations of the state-of-the art}
The limitations of the baseline algorithms for clean speech signal are:
\begin{itemize}
\item Spurious boundaries are introduced in the 
silence part of the speech signal, due to the high 
sensitivity of the STM computation algorithm which is marked as label $S$ in Figure \ref{figure:original_spectrogram}.
\item Over-segmentation in the vowel regions due to the large duration of vowels which marked as label $V$ 
Figure \ref{figure:original_spectrogram}.
\end{itemize}
Apart from clean speech, signal distortion like clipping and noise 
further degrade the performance of the baseline algorithm. 
As shown in Figure \ref{figure:clipping_spectrogram} because of clipping we get spurious boundaries marked as label $C$.
Similarly in Figure \ref{figure:noisy_spectrogram} because of noise we got spurious boundaries marked as label $N$.

\section{Distortions in speech from non-standard recording environments}
\label{sec:Distortions}
Assessment of mis-articulations in 
disordered speech in real-time is done under non-standard environments,
using handheld devices such as mobile phones or tablets. 
However, it was observed that the performance of
algorithms designed for clean speech are degraded
on non-standard recordings speech due to 
(a) clipping - depends on the distance between the user 
and the mobile device.
(b) environment noise
We discuss how the presence of these distortions impacts
phonetic segmentation.
\subsection{The influence of clipping}
\begin{figure}[!t]
\includegraphics[width=0.47\textwidth,height=0.3\textheight]{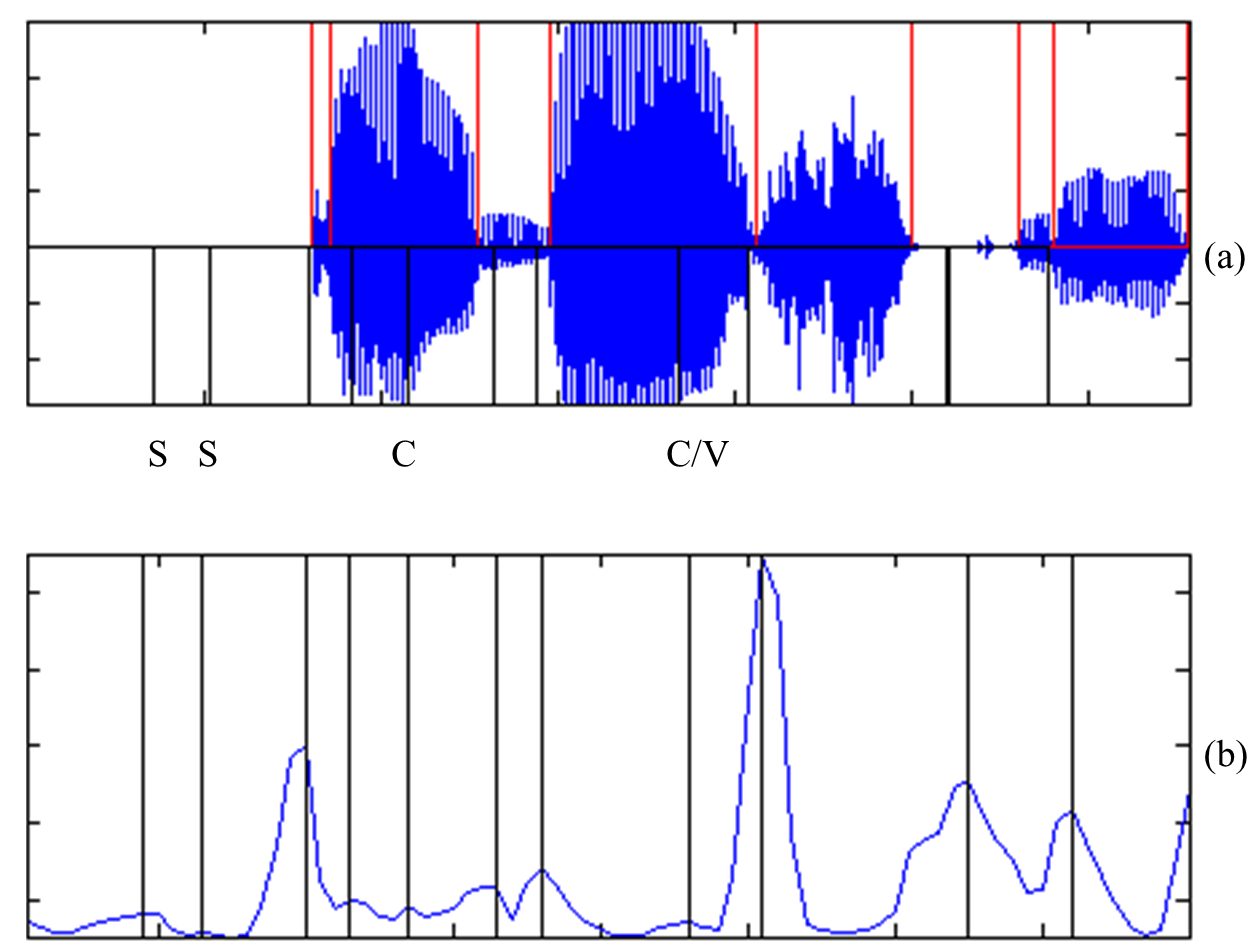}
\caption{{(a) Clipped signal with corresponding manually marked boundaries 
\emph{(red line)} and automatic segmented boundaries \emph{(black lines)} (b) STM contour using baseline algorithm.}}  
\label{figure:clipping_spectrogram}
\end{figure}

Clipping is a distortion that occurs 
when an audio signal level exceeds the dynamic range of 
the recording device.
The only way to avoid clipping is by maintaining the 
recording device at a fixed distance and an uniform volume, which is not
possible in a non-standard environment.
It has been reported in literature 
that presence of clipping in speech reduces the 
speech recognition performance \cite{Harvilla}.
Clipping causes the appearance of 
additional frequencies in the spectral
representation of the speech signal. 
Clearly (see Figure \ref{figure:clipping_spectrogram}),
clipping changes the speech signal and subsequently reflecting in the STM contour. 
Small additional peaks occur in the 
regions of clipped speech of the STM contour.
The proposed post processing technique addresses 
this issue.
\subsection{The influence of environment noise}
\begin{figure}[!t]
\includegraphics[width=0.47\textwidth,height=0.3\textheight]{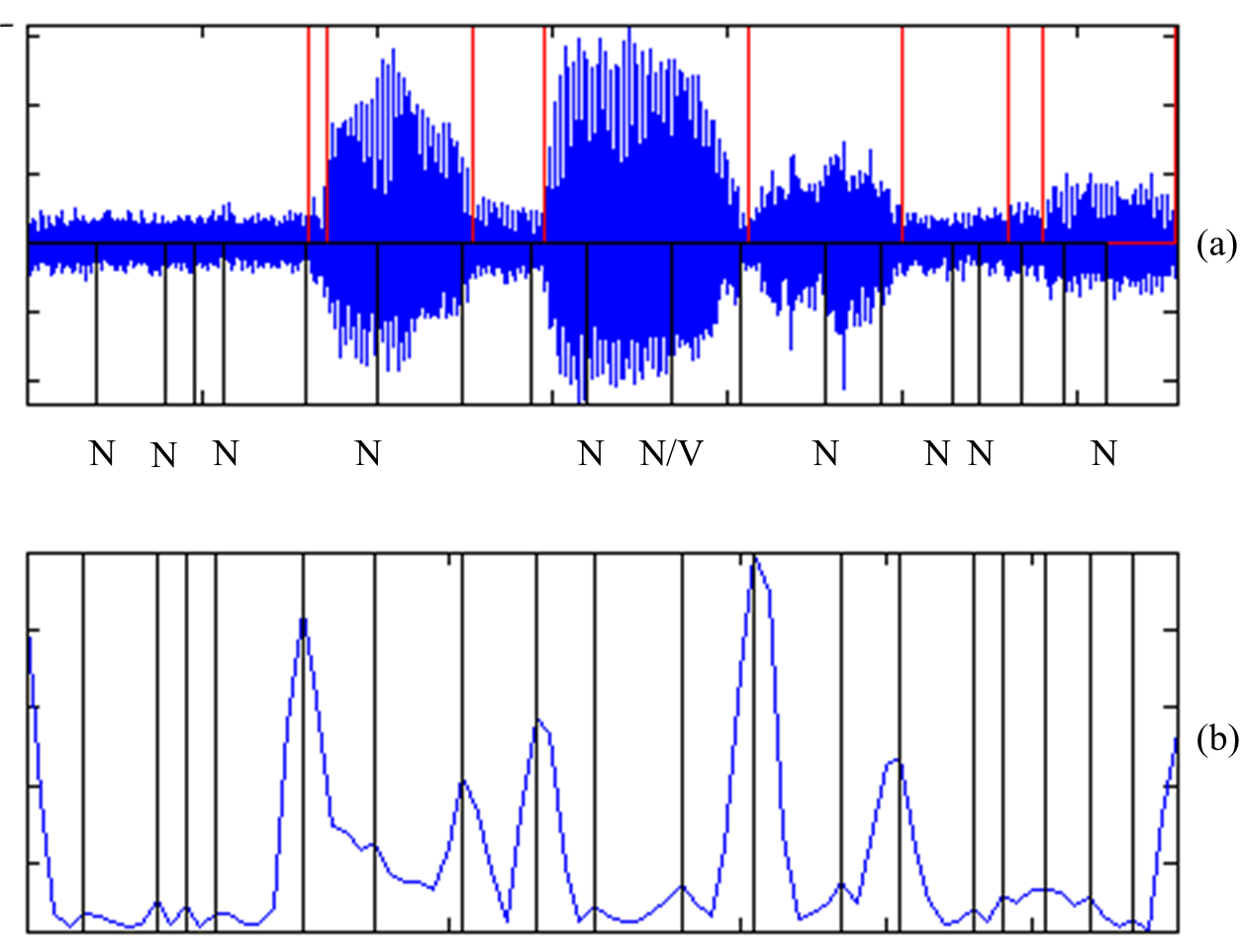}
\caption{{(a) Noisy signal (10 db) with corresponding manually marked boundaries 
\emph{(red line)} and automatic segmented boundaries \emph{(black lines)} (b) STM contour using baseline algorithm.}}  
\label{figure:noisy_spectrogram}
\end{figure}

Noise is an additional unwanted component in the signal. 
It is extremely
difficult even for experts to find out the 
true position of phonetic boundaries in noisy speech signal. 
From Figure \ref{figure:noisy_spectrogram}, it is evident that
presence of noise change speech signal and hence the STM contour.

These distortions have been catered by proposed post processing technique.

\section{Proposed post-processing technique}
This section discusses the novel post-processing method for 
phonetic segmentation. Traditional methods have reported
good accuracy in terms of precision and recall 
for a $20 \thinspace ms$ tolerance. 
However, over-segmentation is a cause of concern when 
applied to our specific application of 
mis-articulation assessment of disordered speech. 
Further, the over-segmentation problem intensifies
in the presence of noise (environmental and clipping).
We propose a post processing mechanism wherein
the STM contour peaks that are considered
as phonetic boundary are selected based on a threshold
determined dynamically for each STM contour unlike
the boundary correction methods that employ experimentally 
determined threshold \cite{Sdusan}.

\mycomment
{In this paper we also propose new method of post-processing the STM contour 
to reduced the number of spurious boundaries as shown in equation \ref{eq:4}. 
By applying proposed post processing techniques it will remove spurious boundaries
from noise like signal from original STM contour. 
This technique will also remove the spurious boundaries inserted in between 
the vowel due to the large duration of the vowel region.}

Median of the STM contour was found 
to be more suitable to be used as a threshold $\tau_M$ for robust phonetic segmentation.

\begin{equation}
STM(m) = 
\begin{cases}
STM(m), & \text{if } STM(m) > \tau_M\\
  \tau_M, &\text{otherwise}
\end{cases}
\label{eq:4}
\end{equation}
The median threshold $\tau_M$, also catered to 
identification and elimination of spurious boundaries inserted within
a vowel due to the large duration of the vowel region. In Section \ref{sec:algo} we will discussed 
the proposed robust phonetic segmentation algorithm.

%
\subsection{Algorithm} 
\label{sec:algo}   
\begin{enumerate}
\item {Read the speech file and corresponding manual transcription}
\item {Extract D = 12 dimension PLPCC feature for each 30 \emph{ms} frame with 20 \emph{ms} overlap}
\item {Compute rate of change of spectral features $\bar{a}$ using Equation (\ref{eq:2}) with I= 2}
\item {Compute mean squared value (STM) using  Equation (\ref{eq:1}) and get the STM contour}
\item {Estimate the meaningful threshold on STM contour}
\item {Modify the STM using Equation (\ref{eq:4})}
\item {Peak peaking on STM contour and estimate the boundaries}
\item {Compare Estimate boundaries and manual boundaries using 20 \emph{ms} tolerance interval}
\item {Compute Precision, Recall and F-score}
\end{enumerate}

In next section we will discussed about experimental setup for testing proposed algorithm.   

\section{Experimental setup}
\label{sec:Experimental setup}
The proposed post processing algorithm was validated on noisy data simulated using clean
TIMIT and Hindi speech corpus. The performance of the algorithm was compared
with (a) clean data (b) with different levels of clipping simulated on clean data
(c) different degrees of noise introduced into the original speech signal.
\subsection{Data Preparation}
\subsubsection{Validation data}
To validate language independence of the proposed algorithm, 
we experimented on two different language databases.
\begin{itemize}
\item TIMIT American English acoustic-phonetic corpus -
This database\cite{Gorofolo} contains utterances from 630 speakers, 
each reading 10 sentences. 
This entire database contains 2,34,925 between-phone boundaries manually determined by experts.
These boundaries do not include the boundaries placed at the beginning and end of the sentence.
\item Hindi acoustic-phonetic speech corpus - 
This speech corpus \cite{CSRLdata} contains a total of $1000$ 
Hindi sentences along with their phone level transcription.
Speech was recorded  as 16 bit PCM, mono at $8 \thinspace kHz$.
10 sentences were spoken by each of the 100 speakers from 11 major 
linguistic regions of India. This entire database contains 55,104 between-phone boundaries manually marked.
\end{itemize}
\vspace{-5mm}
\subsubsection{Test data}
\begin{figure}[htb]
\centering
\includegraphics[width=0.40\textwidth,height=0.40\textheight]{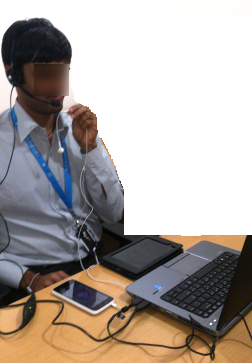}
\caption{{Non-standard recording environment}}  
\label{figure:recording}
\end{figure}

\begin{figure}[!t]
\includegraphics[width=0.50\textwidth,height=0.30\textheight]{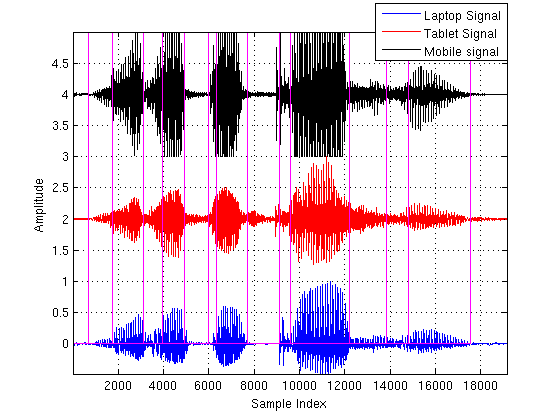}
\caption{Wave file for three different recording medium with corresponding manual segment boundaries.}  
\label{figure:three-file}
\end{figure}

The approach was also tested on data recorded in Hindi on 
three different devices simultaneously. 18 sentences
from 7 speakers was recorded using a setup as shown in the 
Figure \ref{figure:recording}.
Speech on laptop was recorded using a close talking microphone, 
on mobile phone, a hands-free microphone was held in the hand
by the user and the tablet was placed stationarily on the table.
Speech was recorded in wave format at $16 \thinspace kHz$.
We consider the laptop recording as clean data in this set.
Figure \ref{figure:three-file} shows time domain speech signal recorded on 
three different devices laptop, tablet and mobile with corresponding manually marked phone boundaries.
The recorded test data having very less silence/pause region as compare to validation database.
In addition to this, is observed from Figure \ref{figure:three-file} that
laptop recorded signal is clean without any distortion, mobile recorded signal is clipped and tablet 
recorded signal having background noise. This data contains 3997 phonetic segments for each recording device. 
\mycomment{
The text material
consisted of  3 types of sentences.
\begin{itemize}
\item 2 dialect sentences - meant to expose 
the dialectal variants of the speakers and were read by all 100 speakers. 
\item 700 phonetically-rich sentences -designed to provide 
a good coverage of pairs of phones. Each speaker read 7 of these sentences.
\item 100 phonetically-diverse sentences - selected so 
as to add diversity in sentence types and phonetic contexts. 
Each speaker read 1 of these sentences, 
with each sentence being read only by a single speaker.
\end{itemize}
}
\subsection{Simulated data for validation}
We simulate clipping using following transformation.
\begin{equation}
x_c(n) = 
\begin{cases}
x(n), & \text{if } x(n) < \tau\\
\tau. sgn(x[n]), &\text{if } x(n) \geq \tau
\end{cases}
\end{equation}\\
 where $x_c(n)$ is clipped signal of original signal $x(n)$  and $\tau$ is decided 
 based on amount of clipping percentage we introduced for a given speech signal.
 Clipping percentage was varied from 10 to 90 steps of 20.
For simulated noise standard additive white Gaussian noise with varying SNR level 
from between 0 db to 20 db was used. 
Both the baseline approach and the novel approach were validated on the above data.

\subsection{Performance measures}
To measure the performance of proposed approach, we use standard precision-recall measures. 
\begin{itemize}
\item Precision - It is defined as the score of  total number of boundaries estimated in given 20 \emph{ms} tolerance interval 
to the total number of detected boundaries. 
\item Recall - It is defined as the score of total the number of boundaries estimated in given tolerance interval 
to the total number ground truth boundaries. Recall is same as \%Accuracy. 
\item F-score - It is defined as,
\begin{displaymath} 
F-score = \frac{2 \times Precision \times Recall}{Precision + Recall}
\end{displaymath}
Range of these measure varies from 0\% to 100\%.
\end{itemize}
\section{Experimental Results and analysis}
\label{sec:Evaluation}
\begin{figure}[!ht]
\centering
\includegraphics[width=0.45\textwidth,height=0.30\textheight]{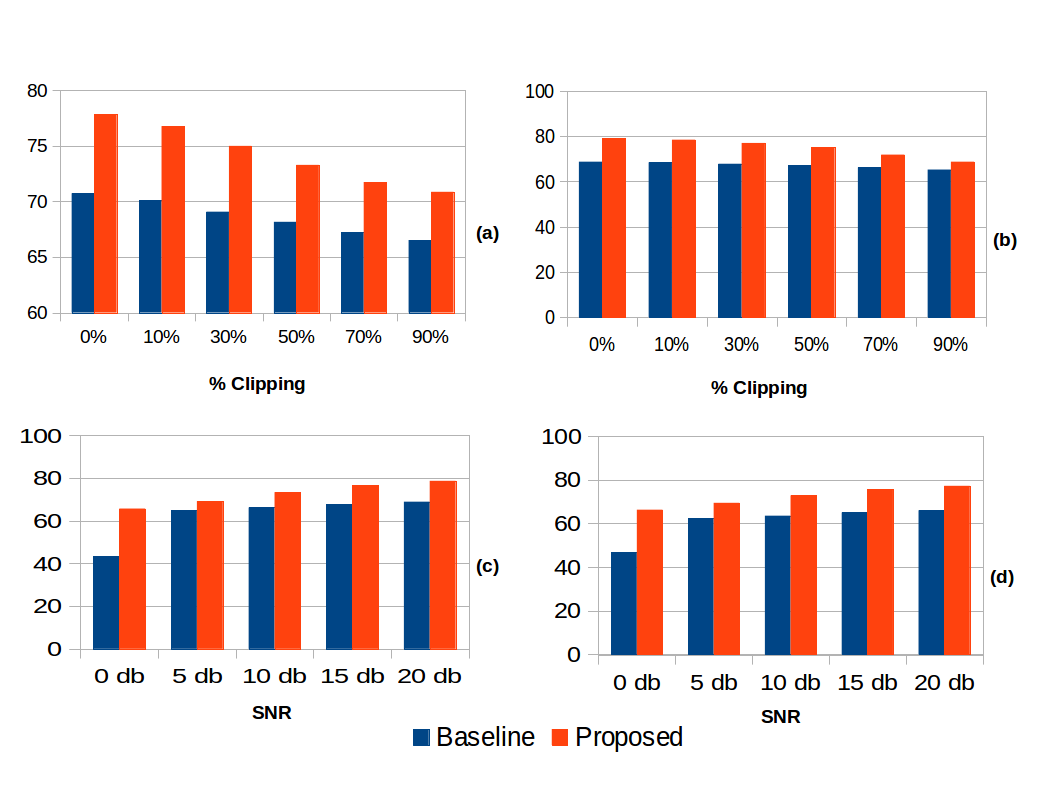}
\caption{(a) \% F-score for TIMIT speech corpus with clipping (b) \% F-score for Hindi speech corpus with clipping
(c) \% F-score for TIMIT speech corpus with noise (d) \% F-score for Hindi speech corpus with noise.}  
\label{figure:results}
\end{figure}

PLPCC speech parameterization was used 
for both baseline and novel approach. A comparison 
of the performance of both the algorithms are as shown in 
Table \ref{table:results-1} for both TIMIT and Hindi speech data.
We got 7\% and 10\% improvement over the baseline algorithm for TIMIT and Hindi data respectively. 
It is evident from the Table \ref{table:results-1} that 
proposed post processing method performs better as compared
to the baseline algorithm.
\begin{table}
\centering
\caption{\% F-score comparison between baseline and proposed techniques for clean speech}
\begin{tabular}{|c|c|c|}
\hline
\textbf{Database}&  \textbf{Hindi} & \textbf{TIMIT}  \\
\hline
\textbf{Baseline} & 68.76 & 70.72\\
\hline
\textbf{Proposed}& 79.22 & 77.84\\
\hline
\end{tabular}
\label{table:results-1}
\end{table}
Figure \ref{figure:results} shows the results of 
proposed post-processing method on simulated clipped and noisy data.
It is evident from the figure that proposed method 
performs better compared to baseline algorithm for distorted data as well.
For up to 50 \% clipping and for SNR greater than 5 \emph(db), 
proposed method gives good results.

Similar experiments were carried out on test data mentioned 
in Section \ref{sec:Experimental setup}. The F-score for 
the three recordings are as shown in the Table \ref{table:results-3} with different tolerance interval. 
\begin{table}[H]
\centering
\caption{\% F-score for speech signals recorded under non-standard conditions using proposed method}
\scalebox{0.69}{
\begin{tabular}{|c|c|c|c|c|c|c|c|}
\hline
\textbf{Tolerance} & \multicolumn{2}{|c|}{\textbf{Laptop}} & \multicolumn{2}{|c|}{\textbf{Mobile}} & \multicolumn{2}{|c|}{\textbf{Tablet}}\\
\hline
\textbf{(\emph{ms})}& \textbf{Baseline} &\textbf{Proposed}&\textbf{Baseline} &\textbf{Proposed} &\textbf{Baseline} &\textbf{Proposed}\\
\hline
\textbf{20}&55.48 &58.25&56.35 & 57.92 & 52.48& 54.09\\
\hline
\textbf{30} &66.12 &69.04 &68.48 &70.04 & 63.97&65.83\\
\hline
\textbf{40} &71.48 &74.34 &75.60 &76.61 &72.38 &73.61\\
\hline
\end{tabular}}
\label{table:results-3}
\end{table}
\section{Conclusion}
\label{sec:conclusion}
Phone level localization of mis-articulation 
is a key requirement for an automatic articulation 
error assessment system. Robust phonetic segmentation techniques
become imperative for such a system. In this work, 
a novel post processing technique was presented for 
phone segmentation under noisy conditions such as 
environment noise and clipping, commonly present
during a mobile phone recording. The proposed approach was 
validated on TIMIT and Hindi speech corpus where the
results show an improvement of 7\% for TIMIT 
and 10\% for Hindi data over the baseline approach.
for both MFCC and PLPCC features, with PLPCC feature 
providing better segmentation. 
Similar  results were seen for a set of speech, recorded simultaneously 
on three devices - a laptop, a stationarily placed tablet 
and a handheld mobile phone, to simulate
different audio qualities in a real-time non-standard
recording environment. 


\bibliographystyle{IEEEbib}
\bibliography{strings,refs}

\begin{thebibliography}{10}

\bibitem{YuanRLSMW13}
Jiahong Yuan, Neville Ryant, Mark Liberman, Andreas Stolcke, Vikramjit Mitra,
  and Wen Wang,
\newblock ``Automatic phonetic segmentation using boundary models.,''
\newblock in {\em INTERSPEECH}, Frédéric Bimbot, Christophe Cerisara, Cécile
  Fougeron, Guillaume Gravier, Lori Lamel, François Pellegrino, and Pascal
  Perrier, Eds. 2013, pp. 2306--2310, ISCA.

\bibitem{Adell04towardsphone}
Jordi Adell and Antonio Bonafonte,
\newblock ``Towards phone segmentation for concatenative speech synthesis,''
\newblock in {\em IN PROCEEDINGS OF THE 5TH ISCA SPEECH SYNTHESIS WORKSHOP},
  2004, pp. 139--144.

\bibitem{PatilJR09}
Vaishali Patil, Shrikant Joshi, and Preeti Rao,
\newblock ``Improving the robustness of phonetic segmentation to accent and
  style variation with a two-staged approach.,''
\newblock in {\em INTERSPEECH}, 2009, pp. 2543--2546.

\bibitem{Stolcke}
Andreas Stolcke, Neville Ryant, Vikramjit Mitra, Jiahong Yuan, Wen Wang, and
  Mark Liberman,
\newblock ``Highly accurate phonetic segmentation using boundary correction
  models and system fusion,''
\newblock in {\em {IEEE} International Conference on Acoustics, Speech and
  Signal Processing, {ICASSP} 2014, Florence, Italy, May 4-9, 2014}, 2014, pp.
  5552--5556.

\bibitem{Mporas}
Iosif Mporas, Todor Ganchev, and Nikos Fakotakis,
\newblock ``Phonetic segmentation using multiple speech features,''
\newblock {\em International Journal of Speech Technology}, vol. 11, no. 2, pp.
  73--85, 2008.

\bibitem{Ziolko}
B.~Ziolko, S.~Manandhar, R.C. Wilson, and M.~Ziolko,
\newblock ``Wavelet method of speech segmentation,''
\newblock in {\em Signal Processing Conference, 2006 14th European}, Sept 2006,
  pp. 1--5.

\bibitem{Tan94applyingwavelet}
Beng~T. Tan, Robert Lang, Heiko Schröder, Andrew Spray, and Phillip Dermody,
\newblock ``Applying wavelet analysis to speech segmentation and
  classification,''
\newblock in {\em Wavelet Applications, volume Proc. SPIE 2242}. 1994, pp.
  750--761, SPIE.

\bibitem{Estevan}
P.~Estevan, O.~Scharenborg, and V.~Wan,
\newblock ``Finding maximum margin segments in speech,''
\newblock in {\em ICASSP}, Honolulu, HI, 2007, pp. 937--940.

\bibitem{King}
Sarah King and Mark Hasegawa{-}Johnson,
\newblock ``Accurate speech segmentation by mimicking human auditory
  processing,''
\newblock in {\em {IEEE} International Conference on Acoustics, Speech and
  Signal Processing, {ICASSP} 2013, Vancouver, BC, Canada, May 26-31, 2013},
  2013, pp. 8096--8100.

\bibitem{Sdusan}
S.~Dusan and L.~R. Rabiner,
\newblock ``On the relation between maximum spectral transition positions and
  phone boundaries,''
\newblock in {\em INTERSPEECH}, Pittsburg, PA, USA, 2006, pp. 17--21.

\bibitem{Bhavik}
N.J. Shah, B.B. Vachhani, H.B. Sailor, and H.A. Patil,
\newblock ``Effectiveness of plp-based phonetic segmentation for speech
  synthesis,''
\newblock in {\em Acoustics, Speech and Signal Processing (ICASSP), 2014 IEEE
  International Conference on}, May 2014, pp. 270--274.

\bibitem{SFurai}
S.~Furui,
\newblock ``On the role of spectral transition for speech perception,''
\newblock {\em Journal of the Acoust. Society of America}, vol. 80, no. 4, pp.
  1016--1025, 1986.

\bibitem{Harvilla}
Mark~J. Harvilla and Richard~M. Stern,
\newblock ``Least squares signal declipping for robust speech recognition,''
\newblock in {\em {INTERSPEECH} 2014, 15th Annual Conference of the
  International Speech Communication Association, Singapore, September 14-18,
  2014}, Haizhou Li, Helen~M. Meng, Bin Ma, Engsiong Chng, and Lei Xie, Eds.
  2014, pp. 2073--2077, {ISCA}.

\bibitem{Gorofolo}
J.~S. Garofolo,
\newblock ``Getting started with the darpa timit cd-rom: An acoustic phonetic
  continuous speech database,''
\newblock {\em National Institute of Standards and Technology (NIST)}, 1988.

\bibitem{CSRLdata}
Sunil~Kumar Kopparapu,
\newblock ,'' \url{https://sites.google.com/site/awazyp/data/speechcorpus},
\newblock Viewed June 2015.

\end{thebibliography}

\end{document}